\documentclass[doublecol]{epl2}

\title{Reward and cooperation in the spatial public goods game}
\shorttitle{Reward and cooperation in the spatial public goods game}

\author{Attila Szolnoki\inst{1} \and Matja{\v z} Perc\inst{2}}
\shortauthor{Attila Szolnoki and Matja{\v z} Perc}

\institute{\inst{1}Research Institute for Technical Physics and Materials Science, P.O. Box 49, H-1525 Budapest, Hungary\\
\inst{2}Faculty of Natural Sciences and Mathematics, University of Maribor, Koro{\v s}ka  cesta 160, SI-2000 Maribor, Slovenia}

\pacs{87.23.Ge}{Dynamics of social systems}
\pacs{02.50.Le}{Decision theory and game theory}
\pacs{87.23.Cc}{Population dynamics and ecological pattern formation }

\abstract{The promise of punishment and reward in promoting public cooperation is debatable. While punishment is traditionally considered more successful than reward, the fact that the cost of punishment frequently fails to offset gains from enhanced cooperation has lead some to reconsider reward as the main catalyst behind collaborative efforts. Here we elaborate on the ``stick versus carrot'' dilemma by studying the evolution of cooperation in the spatial public goods game, where besides the traditional cooperators and defectors, rewarding cooperators supplement the array of possible strategies. The latter are willing to reward cooperative actions at a personal cost, thus effectively downgrading pure cooperators to second-order free-riders due to their unwillingness to bear these additional costs. Consequently, we find that defection remains viable, especially if the rewarding is costly. Rewards, however, can promote cooperation, especially if the synergetic effects of cooperation are low. Surprisingly, moderate rewards may promote cooperation better than high rewards, which is due to the spontaneous emergence of cyclic dominance between the three strategies.}

\begin{document}

\maketitle

\section{Introduction}

Sustainable development and intact social stability require collaborative efforts. Although selfishness and competitiveness are an inherent part of human nature, field studies and experiments attest to the fact that humans are willing to cooperate if the conditions are right \cite{sigmund_10}. Failure to do so results in the exploitation of public goods, such as environmental resources or social benefits, by defectors, who in doing so reap benefits on the expense of cooperators. The ``tragedy of the commons'' succinctly describes such a situation \cite{hardin_g_s68}. In pairwise interactions reciprocation can work in favor of cooperation \cite{trivers_qrb71, axelrod_s81, nowak_n05}. If more than two persons are involved, however, to reciprocate becomes challenging and the burden of sustaining cooperation often falls on punishment \cite{fehr_n02, henrich_s06, hauert_s07, gachter_s08}, as reviewed comprehensively in \cite{sigmund_tee07}. The Achilles' heel of punishment is the fact that it is costly, and it is therefore not clear how it emerges and how to stabilize it. Those that contribute to the common good but abstain from punishing wrongdoers are ``second-order free-riders'', who, in the absence of additional incentives aimed at sustaining punishment, prevail and thus eliminate the threat of sanctioning \cite{panchanathan_n04, fehr_n04, fowler_n05b}. For this unfortunate scenario to unravel, it has recently been suggested that punishment should be a coordinated act \cite{boyd_s10}. It has also been shown that the network reciprocity in structured populations alone may be sufficient to solve the second-order free-rider problem \cite{helbing_ploscb10, helbing_njp10}, and pool-punishment has been considered as an alternative to the traditionally employed peer-punishment with remarkable success \cite{sigmund_n10}. Nevertheless, studies critically probing the effectiveness of punishment in sustaining cooperation, for example in conjunction with anti-social punishment \cite{rand_jtb10}, indirect reciprocity \cite{ohtsuki_n09}, or unfair sanctions \cite{fehr_n03}, are an important reminder of open questions still imbuing the subject.

Reward is an established alternative to punishment \cite{sigmund_pnas01, hauert_jtb10}, albeit studied less frequently in the past. While punishment implies paying a cost for another person to incur a cost, rewards obviously incorporate a cost to bear too, but for another person to experience a benefit. The majority of previous studies addressing the ``stick versus carrot'' dilemma concluded that punishment is more effective than reward in sustaining public cooperation \cite{sigmund_tee07}. But as pointed out in a recent paper by Rand \textit{et al.} \cite{rand_s09}, most of these studies disregarded future consequences for today's actions. Indeed, reputation is key \cite{milinski_n02} and represents a precious asset to loose over an act of punishment that may or may not help in reverting the punished individual. Rewarding is in this respect a safer alternative, and as concluded in \cite{rand_s09}, may be as effective as punishment for maintaining public cooperation. Inspired by these experimental findings, we here investigate the impact of reward on the evolution of cooperation in the spatial public goods game by means of an additional third strategy. The so-called rewarding cooperators, \textit{i.e.} cooperators that reward other cooperators, are willing to bear additional costs in order to reward those that contribute to the common good. As by the introduction of costly punishment, the traditional cooperators, \textit{i.e.} those that contribute to the common good but do not reward other cooperators, become second-order free-riders that fiercely challenge the proliferation of rewarding cooperators. We come to interesting and partly counterintuitive conclusions that go well with existing studies on punishment in structured populations \cite{brandt_prsb03, nakamaru_eer05, helbing_ploscb10, helbing_njp10}, as well as supplement the array of other mechanisms, such as voluntary participation \cite{hauert_s02}, social \cite{santos_n08} and group \cite{shi_dm_epl10} diversity, random exploration of strategies \cite{traulsen_pnas09}, or similar additions \cite{wu_t_epl09, yang_hx_pre09, wang_j_pre09, wang_wx_pre10}, that can be associated with the promotion of cooperation in public goods games.

\section{Public goods game with reward}

The public goods game is staged on a square lattice with periodic boundary conditions, whereon initially each player on site $x$ is designated either as a cooperator ($s_x = {\rm C}$), defector ($s_x = {\rm D}$), or rewarding cooperator ($s_x = {\rm RC}$), with equal probability. Players play the game with their $k=4$ neighbors. Accordingly, each individual belongs to five different groups, \textit{i.e.} it is the focal individual of a Moore neighborhood and a member of the Moore neighborhood of its four nearest neighbors.

Using standard parametrization, the two cooperating strategies (C and RC) contribute $1$ to the public good while defectors contribute nothing. The sum of all contributions is multiplied by the factor $r>1$, reflecting the synergetic effects of cooperation, and the resulting amount is subsequently equally shared among the $k+1$ interacting individuals irrespective of their strategies. In addition, here each cooperator (C and RC) receives the reward $\beta/k$ from every rewarding cooperator that is a member of the focal group, and every rewarding cooperator from this group therefore bears an additional cost $\gamma/k$, thus leading to different payoffs of Cs and RCs. Denoting the number of Cs, Ds, and RCs among the $k$ interaction partners by $N_{\rm C}$, $N_{\rm D}$, and $N_{\rm RC}$, respectively, each cooperator gets
\begin{equation}
P_{\rm C}=r(N_{\rm C}+N_{\rm RC}+1)/(k+1) - 1 + \beta (N_{\rm RC})/k\,\,,
\end{equation}
a defector receives
\begin{equation}
P_{\rm D}= r(N_{\rm C}+N_{\rm RC})/(k+1)\,\,,
\end{equation}
while every rewarding cooperator acquires
\begin{equation}
P_{\rm RC}= P_{\rm C} - \gamma (N_{\rm C}+N_{\rm RC})/k\,\,.
\end{equation}
It is worth pointing out that the cost $\gamma$ and reward $\beta$ are not necessarily identical. This is easy to justify with realistic examples. To praise someone hardly costs anything, yet it may do wonders for the recipient. On the other hand, an affectionate spouse can spend a small fortune on a dress for the partner, only to be later ridiculed for bad taste. While not necessarily representative, we believe these two simple examples suffice to justify the introduction of two rather than a single parameter in order to examine the impact of reward thoroughly, with all its subtleties. We also point out that $\beta$ and $\gamma$ are introduced normalized with the number of neighboring players $k$ in each group in order to facilitate comparisons with results obtained on other interaction graphs or by using differently sized groups. Moreover, the values of $\beta$ and $\gamma$ then represent maximally attainable values within each group and the setup is directly comparable with the previously studied punishment model \cite{helbing_ploscb10}. As was reported in \cite{helbing_njp10, szolnoki_pre09c}, here it holds too that the presented results are robust to reasonable variations of the underlying network structure and group size.

After the three strategies on the $L^2$ square lattice are distributed uniformly at random, a random sequential update with the following elementary steps is performed. First, a randomly selected player $x$ plays the public goods game with the $k$ interaction partners of a group $g$, and obtains a payoff $P_x^g$ from all $k+1=5$ groups it belongs to. The overall payoff is thus $P_x = \sum_g P_x^g$. Next, one of the four nearest neighbors of player $x$ is chosen randomly, and its location is denoted by $y$. Player $y$ also acquires its payoff $P_y$ identically as previously player $x$. Finally, player $y$ imitates the strategy of player $x$ with the probability $q=1/\{1+\exp[(P_y-P_x)/K]\}$, where $K$ determines the level of uncertainty by strategy adoptions \cite{szabo_pre98}. Without the loss of generality we set $K=0.5$, implying that better performing players are readily imitated, but it is not impossible to adopt the strategy of a player performing worse. Such errors in judgment can be attributed to mistakes and external influences that affect the evaluation of the opponent. Each full Monte Carlo step involves all players having a chance to adopt a strategy from one of their neighbors once on average. Depending on the typical size of emerging spatial patterns, the linear system size was varied from $L=400-5000$ in order to avoid finite size effects, and the equilibration required up to $10^7$ full Monte Carlo steps (MCS).

\section{Results}

In the absence of reward, cooperators survive only if $r>3.74$ and crowd out defectors completely for $r>5.49$ if using the square lattice as the interaction graph \cite{szolnoki_pre09c}. These can be used as benchmark values for evaluating the impact of reward on the evolution of cooperation in structured populations. Accordingly, we will focus on three different values of the synergy factor $r$, namely $2.0$, $3.5$ and $4.4$, being representative for low, intermediate and high synergetic effects of cooperation, respectively. Hereafter, we will thus systematically examine how different combinations of reward (the benefit the recipient experiences upon being rewarded) $\beta$ and cost (of giving the reward) $\gamma$ affect the survivability of the three strategies on the square lattice.

\begin{figure}
\centerline{\scalebox{0.4}[0.4]{\includegraphics{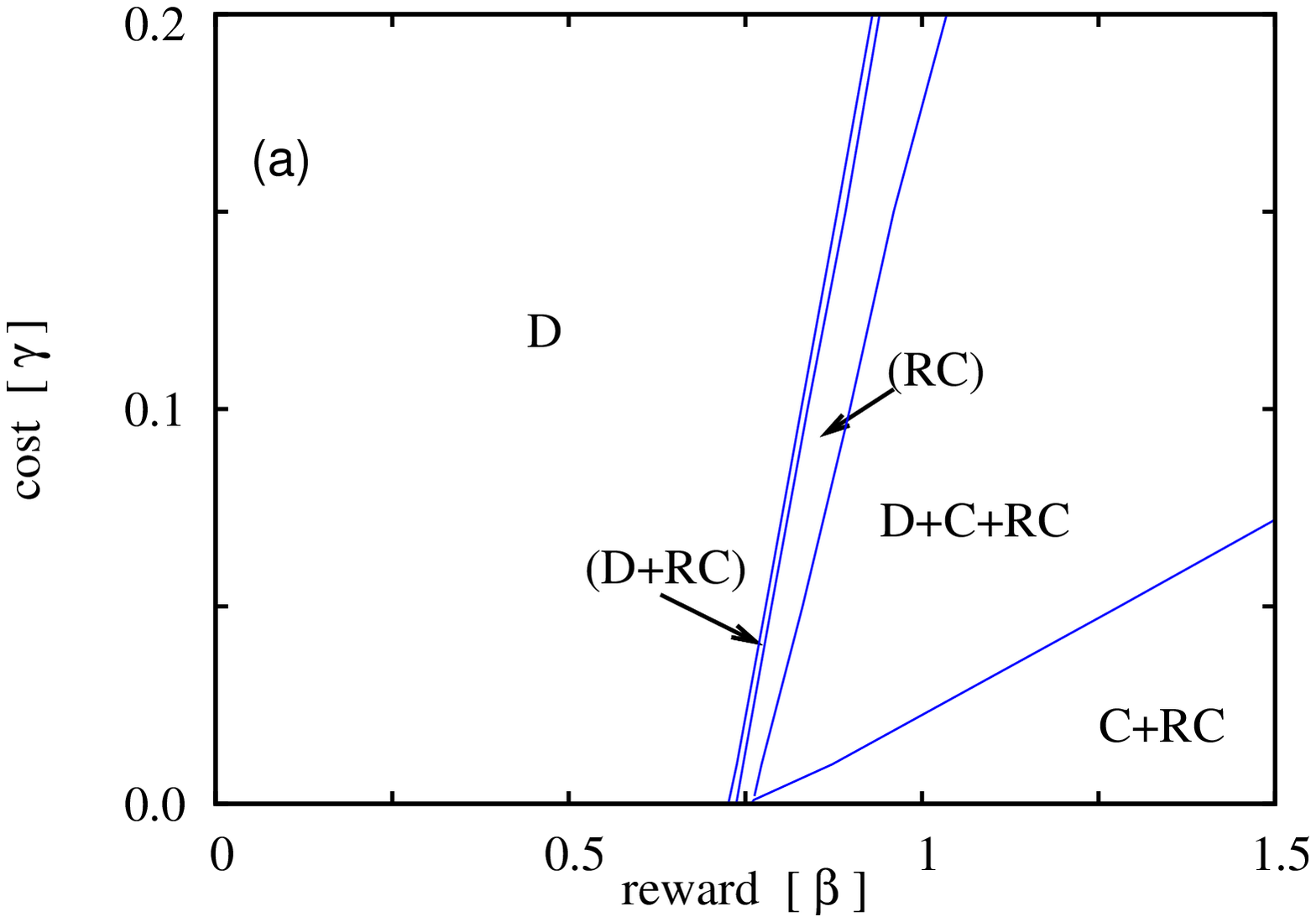}}}
\centerline{\scalebox{0.4}[0.4]{\includegraphics{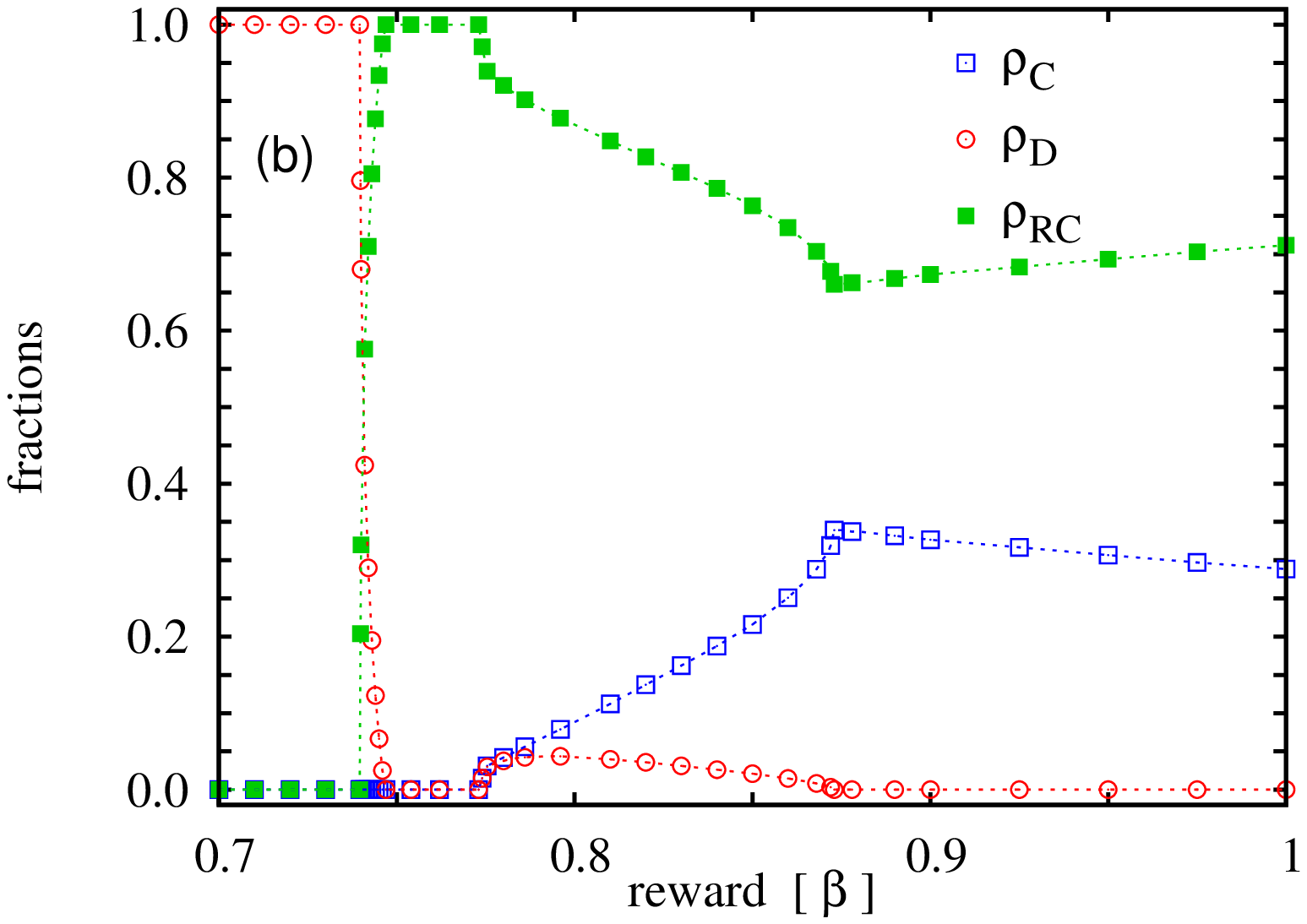}}}
\caption{(a) Full reward-cost phase diagram obtained for the synergy factor $r=2.0$. Different phases are denoted by the symbols of the strategies that survive in the final strategy distribution. Solid blue lines indicate continuous phase transitions. A typical cross-section of the phase diagram at the cost $\gamma=0.01$ is shown in panel (b), depicting the fraction of cooperators $\rho_{\rm C}$, defectors $\rho_{\rm D}$ and rewarding cooperators $\rho_{\rm RC}$ in dependence on the reward $\beta$.}
\label{fig1}
\end{figure}

We start with the low $r$ limit, thus setting $r=2.0$. Figure~\ref{fig1}(a) features the full reward-cost phase diagram, where it can be observed that the pure D phase first gives way to a very narrow region of coexistence of D+RC and shortly thereafter reaches the pure RC phase as the reward increases. The blue transition lines, indicating continuous second-order phase transitions, lean towards higher rewards for larger costs, yet this effect is expected and validates the behavior of the examined model. Most remarkable is the reappearance of defectors in a stable D+C+RC phase if the reward is increased further, thus giving rise to a stable coexistence of all three strategies. Finally, if the reward is higher still and the costs remain moderate (note that the slope of the rightmost transition line is considerably larger), defectors again die out and leave C and RC as the only remaining strategies. Notably, here C and RC are not equivalent strategies as was the case in a recently studied punishment model \cite{helbing_ploscb10, helbing_njp10}, and thus their stable coexistence is possible.

\begin{figure}
\centerline{\scalebox{0.298}[0.298]{\includegraphics{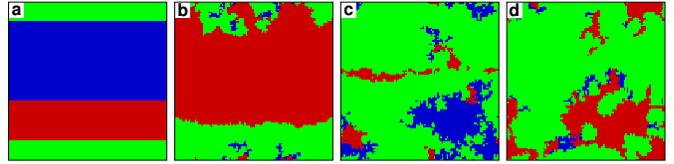}}}
\caption{Characteristic snapshots of a $100 \times 100$ square lattice with specially prepared initial conditions (see main text for details). Colors red, green and blue depict the location of defectors (D), rewarding cooperators (RC) and cooperators (C), respectively. The snapshots were taken at 0 (a), 140 (b), 560 (c) and 600 (d) full MCS, and the parameter values were $r=2.0$, $\gamma=0.05$ and $\beta=0.9$.}
\label{fig2}
\end{figure}

Turning to the reappearance of defectors for intermediate rewards, we show in Fig.~\ref{fig1}(b) a characteristic cross-section of the phase diagram obtained for $\gamma=0.01$. In agreement with the four blue lines depicted in Fig.~\ref{fig1}(a), we can observe four continuous phase transitions. From left to right we have, first, the emergence of rewarding cooperators ($\rho_{\rm RC}>0$), which is quickly followed by the extinction of defectors ($\rho_{\rm D} = 0$). Subsequently, defectors (D) reaper with pure cooperators (C) to form the coexistence of all three strategies, and finally, at $\beta \approx 0.873$ defectors die out again. Interpreting these observations, for sufficiently large $\beta$ the rewarding cooperators can support each other and protect themselves against the invasion of defectors. In accordance with the well-known network reciprocity mechanism, rewarding cooperators aggregate into compact clusters with a smooth interface (not shown here). At still higher $\beta$, the efficiency of rewarding cooperators is so strong that defectors cannot survive. Remarkably, for $\beta>0.775$ the support of cooperative actions becomes powerful enough to enable not just the proliferation of rewarding cooperators (RC), but also the survivability of pure cooperators (C). But since the synergy factor is low (r=2.0), the pure cooperators are susceptible to exploitation by defectors and can therefore survive only in the vicinity of rewarding cooperators. Nevertheless, the emergence of pure cooperators simultaneously enables also the survivability of defectors via a stable D+C+RC phase that is governed by cyclic dominance.

The workings of this cyclic dominance can be demonstrated by examining the snapshots of strategy distributions. Figure~\ref{fig2}(a) depicts a prepared initial state, whereafter the movements of the boundaries that separate the three strategies give vital insight into the dominance between them. Due to the small synergy factor $r$, the defectors (red) can easily invade the blue region of pure cooperators. Simultaneously, since the reward is large, rewarding cooperators (green) can outperform defectors. In the midst of rewarding cooperators, however, pure cooperators (blue) can spread as well because they enjoy the significant benefits of reward but do not bear any costs. But as soon as some of the pure cooperators depart from the safe haven of rewarding cooperators, the whole circle of invasion starts anew, leading to an uprise of defectors (red), who are then again conquered by rewarding cooperators, who then again foster the spreading of pure cooperators, and so on. Clearly thus, the three strategies form a closed loop of dominance, which can be observed nicely if following the snapshots presented in Fig.~\ref{fig2} from left to right (qualitatively identical spatial patterns emerge from random initial conditions if the system size is sufficiently large). It is worth emphasizing that if one of the three strategies dies out by chance due to a small system size, the balance within the closed loop of dominance is broken, and accordingly, one of the remaining two strategies spreads across the whole grid. To avoid this, it is therefore paramount to use sufficiently large system sizes. Interestingly, the stationary density of defectors is considerable, but the increase of the $\rho_{\rm D} (\beta)$ function is the more dramatic the larger the cost of reward $\gamma$. This is in agreement with the behavior of predator-prey systems where the direct support of prey will ultimately be beneficial for the predators. Naturally, if the reward $\beta$ is even larger, defectors cannot survive and the system arrives to the mixed C+RC phase, as depicted in Figs.~\ref{fig1}(a) and (b). Note that the qualitative behavior thereafter does not change and the fraction of cooperators and rewarding cooperators converges to a nonzero value. This, however, is a unique consequence of the spatial structure since in well-mixed populations cooperators (C), \textit{i.e.} second-order free-riders, clearly perform better than rewarding cooperators (RC) and should thus become dominant. In fact, the mechanism that allows rewarding cooperators to survive in the sea of second-order free-riders is identical to the one revealed by Nowak and May when studying the two-strategy spatial prisoner's dilemma game on the square lattice \cite{nowak_n92b}. In our case RCs also form compact clusters that allow them to survive the competition with the superior pure cooperators.

\begin{figure}
\centerline{\scalebox{0.4}[0.4]{\includegraphics{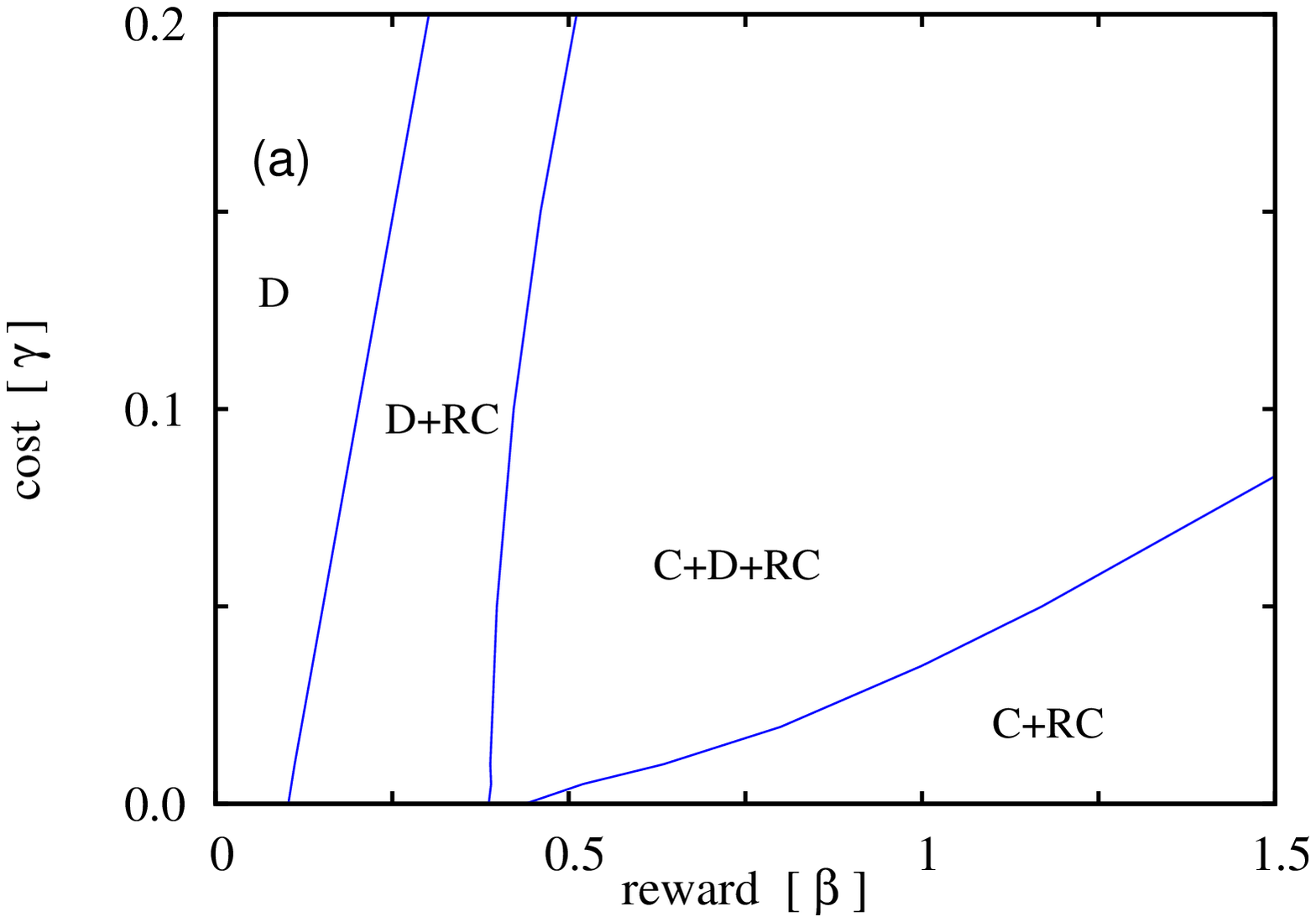}}}
\centerline{\scalebox{0.4}[0.4]{\includegraphics{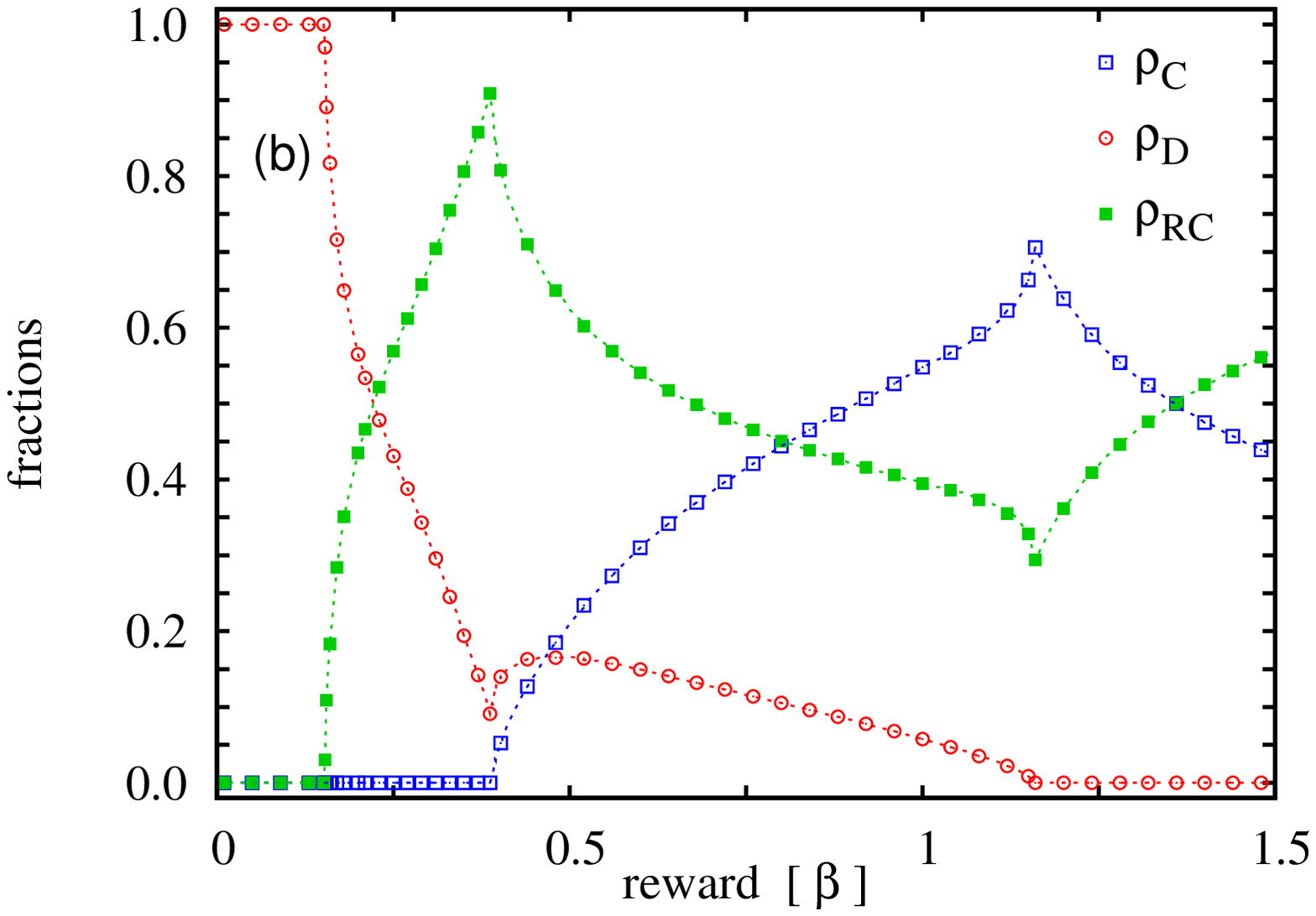}}}
\caption{(a) Full reward-cost phase diagram obtained for the synergy factor $r=3.5$ (phases are denoted by the symbols of surviving strategies). Solid blue lines indicate continuous phase transitions. A typical cross-section of the phase diagram at the cost $\gamma=0.05$ is shown in panel (b), depicting the fraction of cooperators $\rho_{\rm C}$, defectors $\rho_{\rm D}$ and rewarding cooperators $\rho_{\rm RC}$ in dependence on the reward $\beta$.}
\label{fig3}
\end{figure}

To explore the robustness of our observations obtained for the small synergy factor $r=2.0$, we study the evolution of cooperation also for the intermediate value $r=3.5$, although this still results in a pure D phase in the absence of reward \cite{szolnoki_pre09c}. From the reward-cost phase diagram presented in Fig.~\ref{fig3}(a) it follows that the qualitative features, if compared to Fig.~\ref{fig1}(a), remain largely intact. The most significant change is the expansion of the mixed D+RC phase, ultimately leading to the disappearance of the pure RC phase. Note that the stable coexistence of RCs in the sea of Ds is, similarly as the C+RC phase, due to spatial reciprocity \cite{nowak_n92b}, allowing the inferior strategy (as obtained for well-mixed populations) to survive by means of clustering. On the other hand, the coexistence phase containing all three strategies, along with the cyclic dominance between them, is fully preserved. In fact, the D+C+RC region is expanded too, which is further counterintuitive in view of the larger synergy factor used. The latter, of course, promotes cooperation, and should thus act detrimental rather than positive on the survivability of defectors.

\begin{figure}
\centerline{\scalebox{0.4}[0.4]{\includegraphics{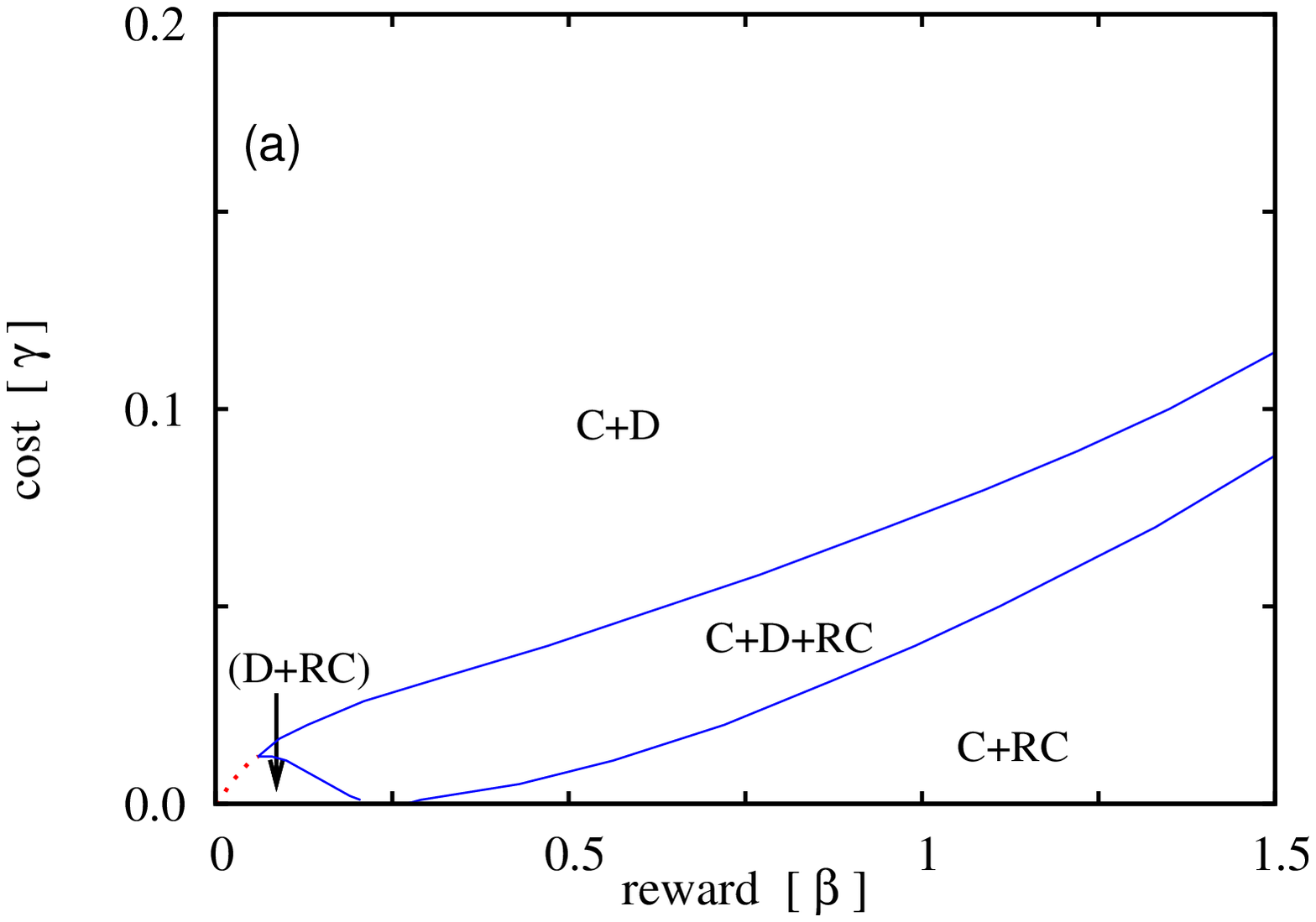}}}
\centerline{\scalebox{0.4}[0.4]{\includegraphics{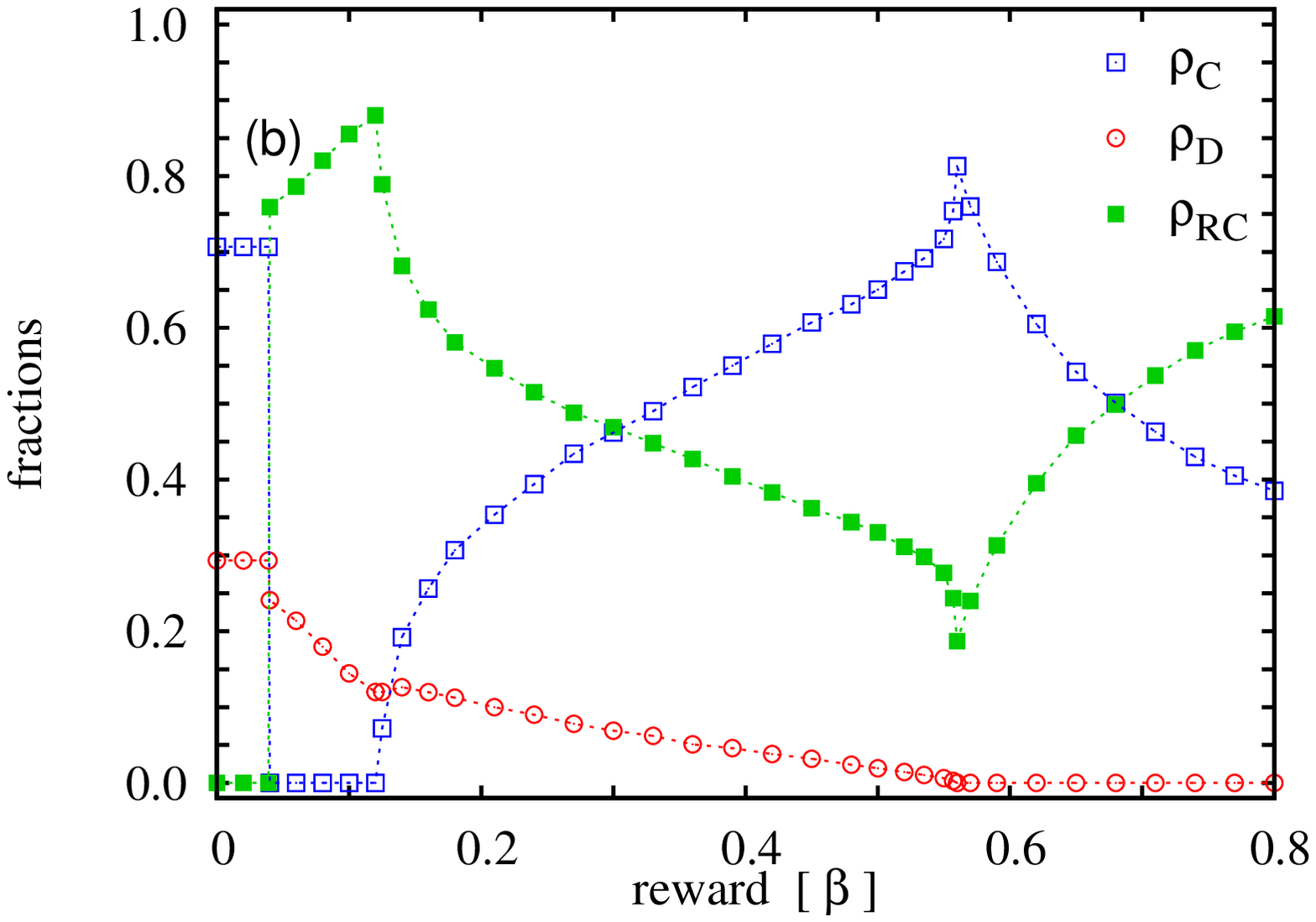}}}
\caption{(a) Full reward-cost phase diagram obtained for the synergy factor $r=4.4$ (phases are denoted by the symbols of surviving strategies). Solid blue (dotted red) lines indicate continuous (discontinuous) phase transitions. A typical cross-section of the phase diagram at the cost $\gamma=0.01$ is shown in panel (b), depicting the fraction of cooperators $\rho_{\rm C}$, defectors $\rho_{\rm D}$ and rewarding cooperators $\rho_{\rm RC}$ in dependence on $\beta$.}
\label{fig4}
\end{figure}

Figure~\ref{fig3}(b) supports this surprising outcome from a quantitative perspective. Indeed, the uprise of defectors, going up to $\rho_{\rm D} \cong 0.19$, is significantly stronger than what was observed for $r=2.0$ in Fig.~\ref{fig1}(b). The reason for this, we argue, is the fact that larger values of $r$ support both, the rewarding as well as pure cooperators. The larger abundance of pure cooperators in particular, gives the defectors more opportunities to conquer lost ground from rewarding cooperators. Note that in the absence of reward $r=3.5$ still fails to sustain cooperative behavior. Accordingly, the strength of dominance within the closed ${\rm D} \to {\rm C} \to {\rm RC} \to {\rm D}$ loop unexpectedly shifts in favor of defectors, which is again an exemplification of how the support of prey ultimately benefits the predator. Furthermore, although not surprisingly, it can be observed that the transition lines in Fig.~\ref{fig3}(a) and the corresponding phase transitions in Fig.~\ref{fig3}(b) are altogether shifted to significantly lower values of $\beta$, which is expected since the synergy factor alone provides a better support for the two cooperative strategies. The emergence of rewarding cooperators, and subsequently also of pure cooperators and defectors by means of cyclic dominance, can thus be warranted already by substantially lower rewards.

Lastly, we examine the impact of reward on the evolution of cooperation at a high synergy factor, thus setting $r=4.4$. The reward-cost phase diagram is presented in Fig.~\ref{fig4}(a). It differs considerably from the previous two, predominantly due to the fact that cooperators can, at this values of $r$, be sustained by network reciprocity alone. Accordingly, the pure $D$ phase is missing and the $\rho_{\rm D} (\beta)$ function is monotonously decreasing, as can be observed from Fig.~\ref{fig4}(b). The existence of the three-strategy phase is also constrained to a significantly smaller portion of the $\beta - \gamma$ parameter plane. Substantial benefits of collaborative efforts thus work clearly in favor of the two cooperative strategies, which become the main aspirants for supremacy on the spatial grid. The perseverance of defectors, going extinct only if $\beta>0.58$, is nevertheless remarkable.

\begin{figure}
\centerline{\scalebox{0.45}[0.45]{\includegraphics{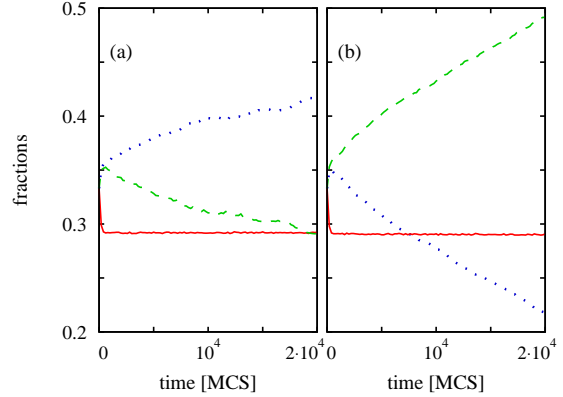}}}
\caption{Time evolution of strategy densities as obtained for $r=4.4$, $\gamma=0.001$, $\beta=0.003$ (a) and $\beta=0.004$ (b). The fraction of defectors is plotted solid red, while the fractions of pure and rewarding cooperators are depicted by dotted blue and dashed green lines, respectively. Note the opposite time evolution of the two cooperative strategies that is induced by a minute change in the hight of the reward $\beta$, taking the system from one side of a discontinuous phase transition to the other.}
\label{fig5}
\end{figure}

Results presented in Fig.~\ref{fig4}(b) allow for an accurate examination of the competition between pure (C) and rewarding (RC) cooperators. Unlike for small and intermediate values of $r$, we can here observe a discontinuous phase transition [marked dotted red in Fig.~\ref{fig4}(a)], by means of which rewarding cooperators first outperform pure cooperators. The mechanism behind this transition is identical to the one reported recently in the context of punishment in structured populations \cite{helbing_ploscb10}, and can be summarized by an indirect territorial battle as follows. Pure and rewarding cooperators form homogeneous isolated islands on the spatial grid and fight independently against the defectors. If the reward is high enough the rewarding cooperators will be more successful in this than pure cooperators, and accordingly will have an evolutionary advantage in the stationary state. Conversely, for less favorable reward conditions, \textit{i.e.} if $\beta$ becomes comparable to $\gamma$, pure cooperators will be more successful in gaining ground from defectors, and accordingly, they will prevail.

These two different evolutionary scenarios can be illustrated nicely by comparing the time courses of strategy densities for two different values of the reward, at one and the other side of the transition line, respectively. As Fig.~\ref{fig5} shows, the fraction of defectors becomes time-independent after a short transient and indeed depends only on the value of $r$. The indirect battle between pure and rewarding cooperators starts thereafter and the fractions of these two strategies will change oppositely depending on $\beta$. If the reward is low, as shown in Fig.~\ref{fig5}(a), pure cooperators outperform defectors more efficiently and hence crowd out also the rewarding cooperators. At the other side of the discontinuous phase transition point, for a slightly higher value of $\beta$, as shown in Fig.~\ref{fig5}(b), the opposite scenario unfolds, and the system will eventually evolve to a D+RC phase. Concluding the study, it can be noted from Fig.~\ref{fig4}(b) that as the reward increases further the second-order free-riders gradually better the rewarding cooperators, for the former enjoy the benefits of reward without participating in sharing the costs. As defectors die out completely the balance shifts again in favor of rewarding cooperators by means of the same mechanism that we outlined when described the results presented in Figs~\ref{fig1}(b) and \ref{fig3}(b).

\section{Summary}

We have investigated the impact of reward on the evolution of cooperation in the spatial public goods game. Using the square lattice as the underlying interaction network, we found that costly rewards facilitate cooperation most effectively if the synergetic effects of cooperation are low. Surprisingly though, high rewards may be less effective in promoting cooperation than moderate rewards. The intricate patterns of cooperation were examined systematically by means of phase diagrams, where a succession of discontinuous and continuous phase transitions was found separating the stable coexistence of different strategies. Depending on the synergy factor and the details of rewarding, we have demonstrated the stable coexistence of all possible combinations of the three strategies. The counterintuitive impact of high rewards in particular, was attributed to the spontaneous emergence of cyclic dominance between the three strategies, which can be molded further by predator-prey-like interactions at intermediate synergy factors. Due to the second-order free-riding role of traditional cooperators who refuse to bear the costs of rewarding, however, it is impossible to conclude that rewards in structured populations render defection maladaptive. Indeed, defection remains viable in considerably large regions of the parameter space, yet even for very low synergy factors, properly tuned rewards can support cooperation where otherwise defection would reign completely. Compared to costly punishment, however, the promotion of cooperation by means of costly rewards seems altogether less efficient. Note that in the absence of defectors the punishing cooperators become equivalent to the cooperators, while rewarding cooperators still keep paying the cost of reward and therefore remain inferior to the second-order free-riders. Thus, for reward to work equally well as punishment, the ratio between the benefit and the cost of rewarding must be significantly higher than in case of punishment (\textit{cf.} \cite{helbing_ploscb10}). At high synergy factors, on the other hand, the network reciprocity alone suffices to decimate the defectors, and the impact of reward is then restricted to establishing the victor between traditional cooperators and rewarding cooperators only. The two duel each other by means of an indirect territorial battle against defectors, where the winning strategy is the one that is more effective in eliminating defectors. If the rewarding is costly the winners are the traditional cooperators, but if the benefits of reward offset its costs by a comfortable margin then the victors are the rewarding cooperators. The border between these two outcomes is a discontinuous phase transition. In sum, the rich plethora of stable pure and mixed phases as well as intriguing dynamical processes that govern the evolution in the presence of rewarding clearly point to the complexity of possible solutions in structured populations and strengthen their prominent role in the pursuit of cooperation.

\acknowledgments
We acknowledge support from the Hungarian National Research Fund (K-73449), the Bolyai Research Scholarship, the Slovenian Research Agency (Z1-2032), and the Slovene-Hungarian bilateral incentive (BI-HU/09-10-001).

\end{document}